\documentclass[fleqn,twoside]{article}
\usepackage{espcrc2}

\usepackage{graphicx}
\newcommand{\bi}{\begin{itemize}}
\newcommand{\ei}{\end{itemize}}

\newcommand{\be}{\begin{equation}}
\newcommand{\ee}{\end{equation}}
\newcommand{\bea}{\begin{eqnarray}}
\newcommand{\eea}{\end{eqnarray}}

\newcommand{\ldm}{\Delta m_{31}^2}

\newcommand{\deltacp}{\delta_{\mathrm{CP}}}
\newcommand{\stheta}{\sin^2 2 \theta_{13}}

\newcommand{\ie}{{\it i.e.}}

\newcommand{\cf}{{\it cf.}}
\newcommand{\etc}{{\it etc.}}

\newcommand{\fig}{Fig.}

\newcommand{\Ref}{Ref.}

\title{Three-flavor analysis of long-baseline experiments}

\author{Walter Winter\address[IAS]{School of Natural Sciences, Institute for Advanced Study,
        Princeton, NJ 08540, USA}%
        \thanks{Work supported by the Leonhard-Lorenz-Stiftung, the SFB 375 of Deutsche
	Forschungsgemeinschaft, and the W.~M.~Keck Foundation},}

\begin{document}

\begin{abstract}
We compare the analysis of existing and future neutrino oscillation long-baseline experiments,
 where we point out that the analysis of future experiments actually implies a 12-dimensional
parameter space. Within the three-flavor neutrino oscillation framework, six of these parameters are the fit parameters, and six are the simulated parameters.
This high-dimensional parameter space requires the condensation of
information and the definition of performance
indicators for the purpose needed. As the most sophisticated example for such an
indicator, we choose the precision of the leptonic
CP phase, and discuss some of the complications of its computation and interpretation.
\vspace{1pc}
\end{abstract}

\maketitle

% LBL, degenerate solutions
Long-baseline experiments, such as conventional beams, superbeams, neutrino
factories, or even new reactor experiments, are just being started with the
K2K accelerator-based experiment~\cite{Nakamura:2000uu}.
Because of the complicated intrinsic structure
of the appearance channels in accelerator-based experiments, correlations~\cite{Huber:2002mx} and
degeneracies~\cite{Burguet-Castell:2001ez,Minakata:2001qm,Fogli:1996pv,Barger:2001yr} play a major role in the analysis of these experiments.
In this talk, we refer to ``correlations'' as connected degenerate solutions
(at the chosen confidence level), and to ``degeneracies'' as disconnected
degenerate solutions (at the chosen confidence level).
The correlations and degeneracies appear in any
{\em fit} manifold, such as a fit to the data.
They come from the intrinsic structure of the
oscillation probabilities, which implies that an experiment cannot uniquely resolve
the individual parameters. In the $\chi^2$-approach, they lead to the final
precision of the quantity of interest
by projection of the fit manifold onto the respective parameter axis (or plane).

% 12-dimensional parameter space
There is, however, one important difference between the analysis of existing and
the simulation of future experiments: For existing experiments, the data are provided by the
experiment, whereas for future experiments, the data have to be simulated.
Thus, the topology of the fit manifold does, for future experiments, not only
depend on the fit parameter values, but also on the simulated parameter values.
The interpretation of such simulated parameter
values is as follows: ``If the actual value, which nature provides, corresponds
to a certain simulated parameter values, then the measurement performance
will be ...''.
Therefore, one actually faces a 6+6-dimensional parameter space if one
simulates a future experiment. Fortunately, the potential simulated parameter
values are not entirely free, since some of them have been already measured
to  certain precisions. For example, the leading solar and atmospheric
parameters are quite precisely known. Nevertheless, it turns out that the
simulated value of $\ldm$ has a rather large impact on the potential of future
accelerator-based experiments. In addition, the simulated values of the mass
hierarchy, $\stheta$, and $\deltacp$ strongly influence the respective
measurements.

% Performance indicators
Because of this high-dimensional parameter space, it is often convenient
to define performance indicators which condense the information. The purpose
of these indicators can be risk minimization with respect to the simulated
parameter values, optimization, or the comparison of different strategies.
Once such indicator is the $\stheta$ sensitivity
limit defined as the largest fit value which fits the simulated value $\stheta = 0$.
Because the simulated value is computed for $\stheta=0$, this sensitivity
limit will not depend on the simulated value of $\deltacp$, as well
as it includes all correlations and degeneracies in a straightforward
way. In addition, it can be shown that it will not
depend on the mass hierarchy~\cite{Huber:2004ug}, either. As performance indicator,
it can serve for the comparison of the potential of different
experiments to extract $\stheta$ from the appearance information,
as well as it can be used for risk minimization with respect to
$\ldm$ or optimization. Similar indicators can be defined for
the mass hierarchy and $\deltacp$, where we will focus on one particular
quantity in the rest of this talk: The precision of the leptonic
CP phase $\deltacp$.

Compared to CP violation measurements, CP precision measurements do not
assume that some values of $\deltacp$ are ``special'', such as the
CP conserving values $0$ or $\pi$, or the maximally CP violating values
$\pi/2$ or $3 \pi/2$. From theory, we know that there has to be some
CP violation at high energies in order to create the
baryon asymmetry. However, there is no evidence that this CP violation is
connected to the low energy leptonic Dirac CP phase, which means
that from theory there is now general argument why certain values of
$\deltacp$ should be realized by nature. Thus, we discuss here the more
 general question of the precision of the measurement of $\deltacp$
 as the most sophisticated measurement in neutrino oscillation physics.
 In particular, for superbeams, where CP violation measurements are very
 difficult, but nevertheless some values of $\deltacp$ might be excluded,
 a new performance indicator is needed. Therefore, we define the ``CP coverage''~\cite{Huber:2002mx} as the range of fit values of $\deltacp$
 which fit a certain true
 value. Hence, a very small CP coverage corresponds to the precision
 of the measurement, whereas a CP coverage close to $360^\circ$ means that
 no information on $\deltacp$ can be obtained. For example, a CP coverage
 of $300^\circ$ implies that no CP violation measurement is possible,
 but $60^\circ$ of all possible values for $\deltacp$ can be excluded.
Eventually, the concept of the CP coverage can be used to evaluate the
performance of next-generation beam
experiments, as well as of future high precision instruments.

CP precision measurements are  strongly influenced
by the simulated values of $\stheta$ and $\deltacp$ itself (besides other parameters).
Two performance indicators are especially useful in this context: The
CP coverage as function of the simulated value of $\stheta$ (``CP scaling'', for fixed
simulated value of $\deltacp$),
and the CP coverage as function of the simulated value of $\deltacp$ (``CP pattern'', for
fixed simulated value of $\stheta$).
The dependence on $\stheta$ (CP scaling) essentially depends (to a first approximation)
on the event numbers in the appearance rates and
is related to the experiment performance. For example, superbeams do have
a different $\stheta$-range, where they return useful results, than neutrino factories
(for an overview, see \Ref~\cite{Winter:2003st}).
Therefore, CP scalings are good performance indicators to compare different classes
of experiments.
The dependence on $\deltacp$ (CP pattern), however, is related to the intrinsic
structure of the
oscillation probabilities and can in simple cases be interpreted in terms of
bi-rate graphs~\cite{Minakata:2001qm,Winter:2003ye}. It can be used for risk minimization
with respect to the unknown true value of $\deltacp$.

The computation of the CP coverage for a fixed given set of simulated parameter
values is, in principle, rather straightforward: For any degeneracy as
well as the best-fit manifold, the fit manifold is projected onto the $\deltacp$-axis.
The CP coverage can then be read off as the fraction of all possible CP values which fit
the chosen simulated value. However, especially for neutrino factories, the computation as well as interpretation of CP patterns and scalings becomes very complicated for several reasons:
\begin{enumerate}
\item
 Neutrino factories have rather good spectral information, which means that there is
 no simple interpretation of the CP patterns in terms of bi-rate graphs.
\item
 The $(\deltacp,\theta_{13})$-degeneracy~\cite{Burguet-Castell:2001ez}
  causes a strong non-Gaussian dependence on the confidence level.
\item
 The $\mathrm{sgn}(\ldm)$-degeneracy~\cite{Minakata:2001qm} starts moving for small values below $\stheta \sim 10^{-3}$ (\cf, \fig~8 of \Ref~\cite{Huber:2002mx}).
\item
 Matter density uncertainties become relevant for large values of $\stheta$~\cite{Burguet-Castell:2001ez,Ohlsson:2003ip}.
\item
 The topology of the fit parameter space becomes very flat below $\stheta \sim 10^{-4}$, \ie,
 it contains lots of local minima at small $\chi^2$-values~\cite{HLWprep}.
\end{enumerate}
Therefore, the CP coverage is not at all a trivial quantity for neutrino factories, since
it contains a lot of highly condensed information. One can even define more condensed performance indicators, such as
a conservative case CP scaling, where one does not choose one fixed simulated value for
$\deltacp$, but computes the conservative case over all values. Such an indicator can
be found in Fig.~19 of \Ref~\cite{Huber:2002mx}. The computation of this indicator, however,
is very sophisticated and requires a lot of computation time.

In summary, the treatment of existing experiments is rather straightforward: One simply
shows all allowed fit regions (degeneracies) to the data. For
future experiments, however, the data have to be simulated, which implies that
the results depend on the simulated parameter values. Therefore, one actually
faces a 12-dimensional parameter space. We conclude that it is hence necessary that one
\begin{itemize}
\item
 identify the most relevant impact simulated parameters.
\item
 identify the purpose of the evaluation (such as optimization, risk
 minimization, comparison of experiments \etc).
\item
 condense the information by the definition of appropriate performance indicators.
\item
 present the performance indicators as function of the relevant impact parameters
 within their currently allowed ranges.
\end{itemize}
One such very sophisticated performance indicator is the CP coverage, which not only
describes the exclusion of different $\deltacp$-values by next-generation experiments,
but also the precision at future high-performance instruments. We have demonstrated several
of the complications of the computation and interpretation of CP coverage
measurements. Especially the analysis of neutrino factories has turned out to be very sophisticated if one wants to identify all of the relevant parameter space.
We conclude that the path leading to future high-precision neutrino oscillation measurements
such as $\beta$-Beams~\cite{Zucchelli:2002sa,Burguet-Castell:2003vv} or neutrino factories~\cite{Geer:1998iz} needs further illumination with respect to the
``complete'' fit and simulated parameter space. This also involves the discussion of conceptual alternatives, such as the ``silver channels''~\cite{Donini:2002rm},
the ``magic baseline'' option~\cite{Huber:2003ak}, or the combination with superbeam upgrades~\cite{Burguet-Castell:2002qx}.

\end{document}